\documentclass{mn2e}

\usepackage{times}
\usepackage{subfigure}
\usepackage{amsmath}
\usepackage{amsfonts}
\usepackage{graphicx}
\usepackage{bm}

\begin{document}
\journal{~}

\title[Model selection optimization]{Optimizing future dark energy
  surveys for model selection goals} \author[C. Watkinson et al.]
      {Catherine Watkinson,$^1$ Andrew
  R. Liddle,$^1$, Pia Mukherjee,$^1$ and David Parkinson$^2$\\ 
$^1$Astronomy Centre, University of Sussex, Brighton BN1
  9QH, United Kingdom\\
$^2$School of Mathematics and Physics, University of Queensland,
  Brisbane, QLD 4072, Australia} \maketitle 
\begin{abstract}
We demonstrate a methodology for optimizing the ability of future dark
energy surveys to answer model selection questions, such as `Is
acceleration due to a cosmological constant or a dynamical dark energy
model?'. Model selection Figures of Merit are defined, exploiting the
Bayes factor, and surveys optimized over their design parameter space
via a Monte Carlo method. As a specific example we apply our methods to generic multi-fibre
baryon acoustic oscillation spectroscopic surveys, comparable to
that proposed for SuMIRe PFS, and present implementations based on the
Savage--Dickey Density Ratio that are both accurate and practical for
use in optimization. It is shown that whilst the optimal surveys using
model selection agree with those found using the Dark Energy Task
Force (DETF) Figure of Merit, they provide better informed flexibility
of survey configuration and an absolute scale for
performance; for example, we find survey configurations with close to optimal model selection performance despite their corresponding DETF Figure of Merit being at only 50\% of its maximum.
This Bayes factor approach allows us
to interpret the survey configurations that will be good enough for
the task at hand, vital especially when wanting to add extra
science goals and in dealing with time restrictions or multiple probes
within the same project. 
\end{abstract}

\section{Introduction}

Cosmology has developed dramatically in recent years; from being
restricted to the realms of philosophy, our observational abilities
have advanced it to the point where we may obtain precise evidence
with which to shape our models and understanding. In this age of
precision cosmology, the fine tuning of surveys can dramatically
improve their performance. This requires us to think in terms of
designer surveys rather than using a build-and-point approach.

For any given problem in cosmology (we use that of dark energy
hereafter) many surveys of varying capabilities will be proposed, and a
combination will make it through the conceptual stages to see the
light of night. For example, there are several stages defined by the
Dark Energy Task Force (DETF) to classify dark energy surveys; stage II
surveys are complete, e.g.\ the Sloan Digital Sky Survey (SDSS);
several stage III surveys are now taking data, e.g.\ The Baryon
Oscillation Spectroscopic Survey (BOSS) and WiggleZ, with others at the
manufacturing stage, e.g.\ The Dark Energy Survey (DES) and the
Hobby--Eberly Telescope Dark Energy EXperiment (HETDEX); and Stage IV
surveys are still in the design phase e.g.\ BigBOSS, the Square
Kilometre Array (SKA), the Wide-Field Infrared Survey Telescope
(WFIRST), and the recently-approved Euclid satellite mission.

When considering the large investments of time,
money and expertise involved in these projects, it is imperative that
designers identify the survey configuration that maximises the science
return. Given the number of surveys all targeting the same goal, it is
also important that they identify the appropriate niche; doing so 
maximises the overall science return from the combined effort of all
relevant surveys. Moreover, naive optimization can be wasteful unless there is an absolute scale of performance that can be used to determine survey configurations that are good enough for a given task, especially when dealing with time or cost restrictions, or with multiple probes within a survey. The importance of optimization for dark energy
surveys was first stressed by Bassett (2005) and Bassett, Parkinson \& Nichol (2005b).

The concept of optimization is universal to design, regardless of the
product in hand; a scalar rating or Figure of Merit (FoM) is defined,
and the configuration of product variables that optimizes this number
is identified. When it comes to designing a survey's observational
parameters it is usual to exploit Monte Carlo Markov Chain (MCMC)
methods to vary things like survey time, area, exposure time, and
redshift range to identify the extreme of a FoM.

In developing their roadmap for the future, the Dark Energy Task Force
defined a FoM for comparing proposed dark energy
surveys.\footnote{We refer here to the FoM of the original report
\cite{det2006}. A subsequent report \cite{det2009} suggested a
more complicated parameter estimation FoM based on principal
components of $w(a)$, but we do not consider that here as our
intention is to deploy alternative FoMs.} The FoM is based on
parameter estimation, quantifying the errors measured on the
$\Lambda\rm CDM$ values of the dark energy equation of state. This has
subsequently become the standard for the quantification of  dark energy
survey performance. However the question we wish to answer in building
these surveys asks: which of the models we have should be
preferred; and if a single model cannot be selected outright, as is
presently the case with the dark energy problem, which can we
discount?
The DETF approach skips this question and assumes that we already know
the right model, the idea being that if the true parameter values lie outside
of the 2-$\sigma$ error contours then the survey will be well placed to
identify it. Whilst this does not seem an unreasonable presumption it
has not been properly tested. 

Bassett (2005) introduced the Integrated Parameter Space Optimization
(IPSO) design framework to address this, proposing that the FoM be some function of the 1-$\sigma$ marginalized dark energy
covariance matrix. In
this paper we take a similar approach, but instead adopt
FoMs that rate a survey's ability to perform model
selection, thereby directly optimizing for the survey's designed
objective. 

It is highly problematic to use frequentist statistics to deal with
model selection, whereas Bayesian statistics provides the perfect
platform. In particular we employ the Bayes factor, which measures the
increase of belief in one model over another that new data
provides. A downside of parameter estimation ratings is that
their scale is relative, providing no simple interpretation of when a
survey is good enough for the job in hand.  This is unfortunate as it is
vital that money and effort does not get frittered away in making surveys
arbitrarily more powerful, whilst promising no significant advances in our
knowledge. The Bayes factor addresses this issue by providing an absolute
scale; this is a big motivation for considering its use within forecasting and optimization
\cite{muk2006,tro2007,tro2007a}.

In this paper we define our model selection FoMs in
section 2; in section 3 we use a particular ground based dark energy
survey aiming to exploit baryonic acoustic oscillations to
identify practical implementations for each, and also to test their
performance; and in section 4 we exploit these model selection ratings
to optimize the baryon acoustic oscillation survey SuMIRe PFS (Subaru
Measurement of Images and Redshifts Prime Focus Spectrograph) \cite{tak2010,tak2010a}.

It is worth noting that the implementations we investigate are relatively crude
and there exists much room for refinements. However, our main
motivation is to compare the performance of the model selection FoM to that
of parameter estimation FoM; refinements to improve on computational
efficiency have no influence on the outcome and are therefore superfluous
at this point. Furthermore, as all optimizations to date exploit parameter
estimation FoMs, we deem it sensible to identify ways in which these
optimizations can be easily adapted to address their short-fallings.

\section{Model selection optimization}

\subsection{Optimization details}

Our principal goal is to introduce the concept of model selection
optimization to astrophysics, the concept being a very general
one. However, for concreteness we will focus throughout on a realistic
scenario where the idea could be deployed, by considering optimization
of baryon acoustic oscillation (BAO) surveys for dark energy that
could be carried out by large multi-object spectrographs on
eight-metre class telescopes.

This study builds on an optimization study that was carried out by
Parkinson et al.~(2010, henceforth P10).
In brief, the survey modelled by this optimization comprised a
3000-fibre spectrograph mounted on a ground-based 8m optical-infrared
telescope. The specifications used
to model this survey are outlined in Table \ref{tbl:specs}, and a full
description can be found in Bassett, Nichol \& Eisenstein (2005a). 
Ultimately this
project (WFMOS) did not move forward to construction, but the concept
and design lives on in the form of SuMIRe PFS, with the
spectrograph to be mounted on the Subaru telescope. Other proposed spectroscopic BAO surveys include BigBOSS (Schlegel et al. 2011) and DESpec.

\begin{table}
  \centering
      \begin{tabular}{l|l}
	\hline
	Constraint Parameter & Value \\
	\hline
	Total observing time & 1500 hours \\
	Field of view & $1.5^{o}$ diameter \\
	$n_{\rm fibres}$ & 3000 \\
	Aperture & 8m \\
	Fibre diameter  & 1 arcsec \\
	Overhead time between exposures & 10 mins \\
	Minimum exposure time  & 15 mins \\
	Maximum exposure time & 10 hours \\
	Wavelength response & 375 to 1000 nm\\
	Width of redshift slices, $dz$ & 0.05
      \end{tabular}
      \caption{Survey constraints used in modelling the multi-object spectrograph.}
      \label{tbl:specs}
\end{table}

\begin{figure}
  \centering
      \includegraphics[width=0.5\columnwidth]{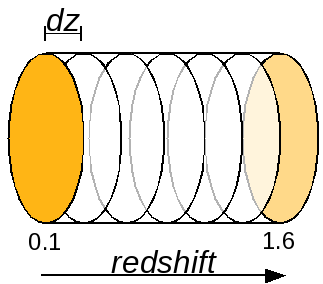}
      \caption{Representation of the $z$ binning method of the
        optimization code used.}
      \label{fig:bins}
  \end{figure}

We model the survey to observe line emission of pre-selected active
star-forming galaxies. Its wavelength coverage allows observation of
the OII lines and the $4000\rm\AA$ break in the redshift range $0.1
\le z \le 1.6$; this overall range is divided into sub-bins as per
Figure \ref{fig:bins} and the density throughout is fixed by that of
the deepest redshift bin.

Our optimization method closely follows that of P10, utilising Monte Carlo Markov Chain (MCMC)
\cite{met1953,has1970} methods to identify the survey configuration
that maximises a FoM. The larger this rating
the better the survey's performance. The variables that describe each
survey configuration are described in Table \ref{tbl:vary}. Given that
we already have dark energy constraints from the Sloan Digital Sky Survey (SDSS) and will soon have data from Planck, SDSS data and forecasts for the Planck data are included as prior information; again refer to P10 for details.

\begin{table}
  \centering
    \begin{tabular}{l|l}
      \hline
      Survey Parameter  & Symbol \\
      \hline
      Time allocated & $\tau$\\
      Area covered & $A$ \\
      Minimum of redshift bin & $z_{\rm min}$ \\
      Maximum of redshift bin & $z_{\rm max}$ \\
      Number of pointings & $n_{\rm p}$ \\
    \end{tabular}
    \caption{Survey parameters varied by the optimization code, affecting the
various FoM under consideration.}
    \label{tbl:vary}
\end{table}

We wish to compare a parameter estimation FoM, that rates a survey's
ability to measure the parameters of interest assuming the true model is known, to model selection FoMs
that recognise our uncertainty surrounding the most preferable model.
The Dark Energy Task Force (DETF) FoM has been widely adopted by the
cosmological community as the standard for comparison of dark energy
surveys and their optimization \cite{det2006}. As such, we take the
DETF FoM as the parameter estimation baseline for comparison.

The DETF FoM makes use of the CPL \cite{che2001,lin2003}
parametrisation  of the dark energy equation of state $w$, given by
\begin{equation}
w(a) = w_0 + w_a\left( \frac{z}{1+z} \right)  \,,
\end{equation}
where $w_0$ is a constant characterising the behaviour of $w$ in the local
universe and the constant $w_a$ characterises its redshift
dependence. The DETF FoM is the inverse of the area confined within
the 95\% confidence level on $w_0$ and $w_a$ measurements, assuming
throughout that $w_0=-1$ and $w_a=0$, i.e.\ $\Lambda\rm CDM$, is the
true cosmological model. The smaller this area, the larger the FoM,
and the more accurate the survey.

For each survey configuration, the optimization of P10 forecasts the errors
on the measurable quantities $d_A$ and $H$ by using a 1D Fisher matrix
based transfer function as derived by Seo \& Eisenstein (2007). This returns the
BAO distance errors as a function of survey properties and
non-linearity. These errors are then translated onto $w_0$ and $w_a$,
given by the inverse of the marginalised Fisher matrix
i.e.\ $F^{-1}_{w_{0}w_{a}}$; from this the DETF FoM may be calculated:
$$
\mathrm{FoM_{DETF}} = \frac{1}{\sqrt{
    \sigma^{2}_{w_{0}w_{0}}\sigma^{2}_{w_{a}w_{a}} -
    \sigma^{4}_{w_{0}w_{a}} }} = \frac{1}{\sqrt{\det
    F^{-1}_{w_{0}w_{a}} } } \,.
$$
For detailed information on this optimization and Fisher matrix
approach please refer to P10 and Parkinson et al.~(2009). 

\subsection{Model selection FoM}\label{sec:modselFoM}

As mentioned we use Bayesian Statistics as the foundation for our
model selection FoMs. As this subject has been covered extensively
in the literature, we will only provide an overview here. The basic laws of probability such as the multiplication rule were
shown by Cox (1946) to be the mathematical framework of Boolean
logic. Bayes theorem derives from direct application of this rule and
provides a means to calculate the probability of a given model ($M$)
(as per equation \ref{eq:modpost}) or hypothesis
($\boldsymbol{\theta}$) in light of data ($D$)
\cite{jef1961,jay2003,mac2003,gre2005}:
\begin{equation}
p(M \mid D) = \frac{p(D \mid M)\,p(M)} {p(D)}\label{eq:modpost} \,.
\end{equation}
Bayesian statistics provide a
natural framework for dealing with model selection and as such form
the basis for the model selection FoM used in this paper.

From here on we use standard notation when referring to probabilities,
e.g.\ $p(A \mid B,C)$ means the probability of $A$ given
that $B$ and $C$ are true. In equation \ref{eq:modpost}, $p(M
\mid D)$ is referred to as the model posterior; $p(D \mid M)$ is the
model likelihood, generally referred to as the evidence; the prior
$p(M)$ characterises our state of knowledge before the data was
collected; and the normalisation term is $p(D)$, the probability of
the data.

The star of the show is the Bayes factor $B$ \cite{jef1961,kas1995},
which measures the increase of belief in one model over another given
new data. Alternatively it can be considered as the change in model
odds from before the data was considered to after. This scalar
quantity is evaluated by taking the ratio of the evidence $E$ of one
model given data, i.e.\ $p(D \mid M_0)$, to that of another, $p(D \mid
M_1)$, as given in equation \ref{eq:bayes}:
\begin{equation}
B = \frac{E(M_0)}{E(M_1)} =\frac{p(D \mid M_{0})}{p(D \mid
  M_{1})} \label{eq:bayes} \,. 
\end{equation}

The combination of model selection FoMs we use for this work takes
account of the uncertainty in our knowledge. We allow the assumed
model to vary through its values of $w_0$ and $w_a$, rather than fixing
it to one fiducial model as is the case for the DETF FoM. The allowed
models are restricted to a chosen region of $w_0$--$w_a$
parameter space, in which $-2 \le w_0 \le -0.33$ and $-1.33 \le w_a
\le 1.33$. This restricted parameter space summarises the prior range used in our calculations.

A plethora of models exists offering explanation for dark energy, see
Caldwell \& Kamionkowski (2009) and references therein. Here, two overarching models
are considered; $\Lambda\rm CDM$ ($M_0$) for which $w_0=-1$ and
$w_a=0$, and evolving dark energy ($M_1$) where $w_0$ and $w_a$ can
have any values chosen uniformly within the confines of the above prior range.
Future observational indications of a deviation from the $\Lambda\rm CDM$ case would no doubt prompt a much wider investigation of both dynamical dark energy models and modified gravity models, but for this work a
two-model approach is sufficient.

We test two Bayes factor based model selection FoMs, first defined by
Mukherjee et al.~(2006a, M06 hereafter), for use in
optimization.\footnote{Alternative model selection FoMs, also suitable
for these purposes, have been given in Trotta (2007a,b) and Trotta et al.~(2011). Unlike those used here, these FoMs average over the present state
of knowledge.} We also speak in terms of $\ln
B$ for the bulk of this paper, whereby even odds translate to $\ln
B=0$, positive values support the simpler $\Lambda$CDM model and negative
values support the more complex model.\footnote{It is equally valid to
assign M$_0$ to be the evolving dark energy model instead of $\Lambda$CDM,
in which case this interpretation is inverted.} As in M06 we make use
of the Jeffreys'
scale \cite{jef1961}, outlined in Table \ref{tbl:jeff}, to judge the
significance of $\ln B$. In constructing our FoM we treat $\ln B$ as a function
of the dark energy model we assume to be `true', that is $\ln B(w_0, w_a)$.

\begin{table}
  \centering
    \begin{tabular}{|l |l|}
      \hline
      $|\ln B|$ range  & Level of significance\\
      \hline
      $|\ln B| < 1$ & Not worth mentioning\\
      $1 < |\ln B| < 2.5$ & Significant\\
      $2.5  < |\ln B| < 5$ & Strong\\
      $5 < |\ln B|$ & Decisive\\
      \hline
    \end{tabular}
    \caption{The Jeffreys' Scale provides a useful guide when
      interpreting the Bayes factor.}
    \label{tbl:jeff}
\end{table}

\begin{figure}
	\centering
		\includegraphics[width=0.9\columnwidth]{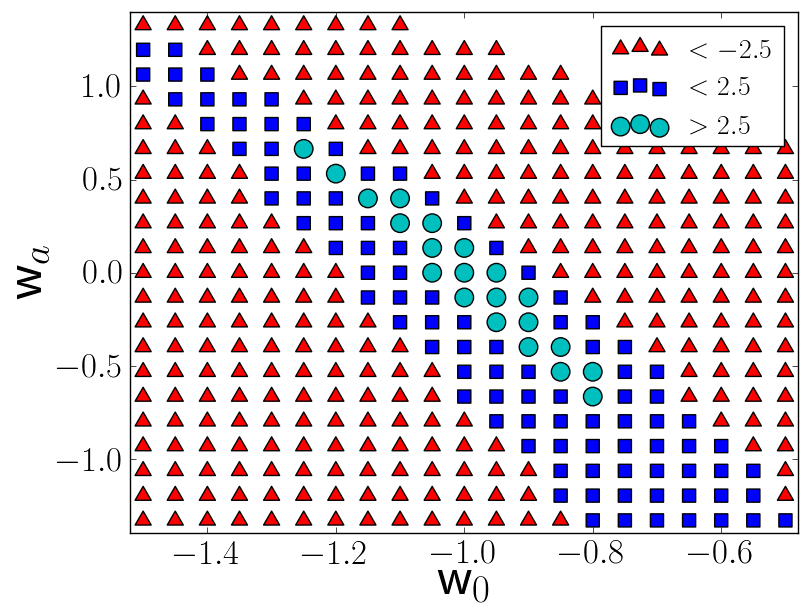} 
	\caption{Each point in this $w_0$--$w_a$ plot is in turn assumed
          to be the true model, and the Bayes factor of $\Lambda$CDM
          versus dark energy is calculated at that point. The (green)
          circles  mark the region where $\Lambda\rm
          CDM$ is preferred, i.e.\ $\ln B>2.5$; (blue) squares mark the
          region in which $\Lambda\rm CDM$ is not discounted,
          i.e.\ $-2.5<\ln B<2.5$; and the (red) triangles correspond to
          the region in which evolving dark energy is correctly
          preferred, i.e.\ $\ln B < -2.5$. }
	\label{fig:DEscatter}
\end{figure}

\begin{enumerate}
\item \textbf{Assuming constant $\boldsymbol{w}$}

The first of these FoMs measures how strongly a survey will support
$\Lambda\rm CDM$ when it is the true underlying model. This is done by
setting $w_0 = -1$ and $w_a=0$ in all calculations contributing to the
Bayes factor forecast. The larger $B$ is at this point in $w_0$--$w_a$
parameter space, the stronger the survey if $\Lambda\rm CDM$ does
transpire to be the true model. This FoM will be referred to as
$\boldsymbol{\ln B(-1,0)}$ hereafter. One of its useful properties is that it gives an absolute scale of support for $\Lambda$CDM; i.e.\ if future experiments continue to increase support for this paradigm, it gives a criterion by which we can decide whether we have done enough to satisfy ourselves, and should turn to other scientific questions.

\item \textbf{Assuming evolving $\boldsymbol{w(z)}$} 

The second model selection FoM measures a survey's ability to discount
$\Lambda\rm CDM$ when it is not the true model. To evaluate this we
forecast the Bayes factor as a function of $w_0$ and $w_a$ and
calculate the area of $w_0$--$w_a$ parameter space in which the survey
will not be able to discount $\Lambda\rm CDM$. This is done by finely
gridding this parameter space and for each point on the grid
forecasting the Bayes factor assuming the $w_0$ and $w_a$ values at
that point.

The FoM that we will refer to as \textbf{area}$^{\boldsymbol{-1}}$
hereafter is the inverse of the area containing values of $\ln B >
-2.5$, corresponding to the green circles and blue squares of Figure
\ref{fig:DEscatter}. The larger this figure the more powerful the
survey for constraining evolving dark energy models and the greater
the chance of detection of evolution if present. Unlike $\ln B(-1,0)$ which
has a direct probabilistic interpretation, the area$^{-1}$ does not; it 
instead provides a measure of a survey's predicted interpretation of
dark energy model-space.

\end{enumerate}

\subsection{Methods for calculating the Bayes factor}

Two different approaches are utilised in calculating $\ln B(w_0,w_a)$:
nested sampling as devised by
Skilling (2006); and the Savage--Dickey Density Ratio, first introduced
in a cosmological context by Trotta (2007a).

\subsubsection{Nested sampling}

Any model is defined by a set of cosmological parameters
$\boldsymbol\theta$; for example, $\Lambda\rm CDM$ can be described by
$\boldsymbol\theta = [w_{\rm de}, \Omega_{\rm de}, \Omega_{\rm m},
  \Omega_{\rm r}, \Omega_{k}, \Omega_{\rm b}, H_{0}, n_{s}, \sigma_{8}
]$. The values of each of these parameters must be estimated by means
of best fit to the data. This can be done using Bayes theorem, as per
equation \ref{eq:bestfit},
\begin{equation}
p(\boldsymbol\theta \mid D, M) = \frac{p(D \mid \boldsymbol\theta,
  M)p(\boldsymbol\theta\mid M)} {p(D \mid M)} \,,\label{eq:bestfit} 
\end{equation}
with MCMC methods identifying the point
in cosmological parameter space at which the posterior
$p(\boldsymbol\theta \mid D, M)$ is maximised \cite{hob2010}.

Notice the denominator in equation \ref{eq:bestfit} is the evidence
required for the Bayes factor of equation \ref{eq:bayes}. By
integrating over all allowed values of this parameter set
$\boldsymbol{\theta}$ it is possible to calculate the evidence using
equation \ref{eq:evi}; the evidence is therefore the average
likelihood over the prior parameter space, thus rewarding models for
predictive power.

\begin{equation}
E(M) = p(D \mid M) = \int p(D \mid \boldsymbol{\theta}, M)
p(\boldsymbol{\theta} \mid M) d\boldsymbol{\theta} \label{eq:evi} \,. 
\end{equation}

Nested sampling \cite{ski2006} recasts this multi-dimensional evidence
integral in 1D by integrating over the prior mass $X$, where $dX =
p(\boldsymbol{\theta}\mid M)d\boldsymbol{\theta}$ and $L$ refers to
the likelihood:
\begin{equation}
E = \int_{0}^{1}L(X)dX \,.
\end{equation}
The nested sampling algorithm starts by sampling a large number of
points from the likelihood surface simultaneously and assigns equal
fractions of the total remaining prior mass to each sample.  It then
proceeds by adding the lowest probability point ($L_j$) (whose prior
mass is $X_j$) to the evidence integral sum:
\begin{equation}
E = \displaystyle\sum\limits_{j=1}^m \frac{L_j}{2}(X_{j-1} - X_{j+1}) \,.
\end{equation}
The algorithm then reduces the remaining prior mass, by a
statistically estimated amount. The lowest likelihood sample is
replaced with a sample randomly selected from the prior, with the sole
selection criteria that it be of higher likelihood than the
previous. The main challenge in implementing the algorithm is to find
a way to carry out this sampling efficiently, a simple approach being
ellipsoidal sampling \cite{muk2006b} and a more sophisticated approach
suitable for multi-modal likelihoods being to partition the points
into clusters of ellipsoids \cite{fhb}.

This entire process is repeated, building up the evidence sum, until
the accuracy has reached an acceptable level. At the point of
termination the remaining contribution to the evidence integral is
added. As this is a numerical estimation, several repeats are done
from which the mean and error are extracted. A detailed account of the
nested sampling implementation we use is given in Mukherjee et al.~(2006b).

Calculations of the Bayes factor are in principle simple using nested sampling.
The evidence is calculated by first assuming $\Lambda\rm CDM$ as the true model, then independently by assuming evolving dark energy, when simulating survey data. 
Unfortunately, due to the very large number of computations
required to sample both survey and model parameter spaces, nested sampling
is too inefficient to be regarded as practical in full MCMC
optimizations. 
However it is still possible to utilise nested sampling when 
investigating the $\ln B(-1,0)$ FoM for very basic manual optimizations, e.g.\ manually altering only one survey parameter at a time.

\subsubsection{Savage--Dickey Density Ratio (SDDR)}

As the nested sampling algorithm is too slow to be seriously
considered for the full scope of model selection optimization that we wish to
consider, the Savage--Dickey Density Ratio (SDDR) is investigated. The SDDR is a
simplification of the Bayes factor that assumes a less complex model
is nested within a more complex model and that the priors are
separable. For example, $\Lambda\rm CDM$ is nested within the evolving
dark energy model's parameter space where $w_0=-1$ and $w_a=0$, and
furthermore the priors
concerned with these two dark energy parameters ($\boldsymbol{w}$) and
those concerned with the nuisance parameters ($\boldsymbol{N}$) of the
models can be separated,
i.e.\ $p(\boldsymbol{w},\boldsymbol{N})=p(\boldsymbol{w})p(\boldsymbol{N})$.
The SDDR is given by equation \ref{eq:SDDR1},
\begin{equation}
B = \frac{p(\boldsymbol{w} \mid D)} {p(\boldsymbol{w})}
\Bigg|_{\boldsymbol{w}=\boldsymbol{w^*}} \,,
\label{eq:SDDR1}
\end{equation}
where $\boldsymbol{w^*}$
represents the simpler models' nested values, being a special case of the more complex model's parameter vector $\boldsymbol{w}$. For a derivation of this see Appendix B of
Trotta (2007a).

This allows the Bayes factor to be evaluated by considering the
marginalised posterior probability of the more complex model and its
prior at the parameter values of the nested simpler model. This
removes the need for the computationally expensive integral as required
to calculate the evidence via equation \ref{eq:evi}. Both assumptions
made in deriving equation \ref{eq:SDDR1} are true for the dark energy
models under consideration and nothing has been assumed about the
likelihood, therefore it is exact in this case. However, we now make a
further assumption that makes this implementation approximate.

To minimise alteration to the original DETF optimization and hence
calculation time, our SDDR calculation assumes Gaussianity of the
posterior in $w_0$ and $w_a$ having marginalized over all other
parameters. The Bayes factor can therefore be forecast with only a few
simple additions to the DETF optimization, by application of the following:
\begin{equation}
  \centering B(w_{0},w_{1}) = \frac{\Delta w_{0}\Delta
    w_{1}}{2\pi\sqrt{\det F^{-1}}}e^{-\frac{1}{2} \sum_{\mu\nu}(
    w_{\mu}- w^*_{\mu})F_{\mu\nu}( w_{\nu}- w^*_{\nu})} \,.
\label{eq:SDDR2}
\end{equation}
In this equation $\nu$ and $\mu$ can have values of either 0 or 1;
$\boldsymbol{w^{*}}$ are the nested values of the simpler model;
$w_1=w_a$; $\Delta w_{0}$ and $\Delta w_{1}$ are the width of the flat
prior ranges; and $F_{\mu\nu}$ is the marginalised Fisher matrix. A numerical approach using finite-differencing was used to determine the Fisher matrix, the details of which are described in Appendix A of P10.

Recall that values larger than unity ($\ln B > 0$) support the simpler
model, and values less than unity ($\ln B < 0$) support the more
complex model. We see then that the pre-factor of equation \ref{eq:SDDR2}
acts as an
amplitude, measuring the ratio of the area of $w_0$--$w_a$ parameter
space allowed by the more complex model to the area of the error
ellipse; this term therefore penalises the more complex model for
unjustified parameter space. The exponential part measures the
distance between the two models and can lend support for the more
complex model by suppressing the amplitude term.

This SDDR approach allows investigation of both model selection
FoMs. However for the area$^{-1}$ FoM to be practical for full MCMC
optimizations it needs refining; for example, $\ln B(w_0,w_a)$
calculations could be parallellized and MCMC could be used
to determine the area$^{-1}$.

\subsubsection{A note on priors}

The prior range mentioned in section \ref{sec:modselFoM} is the same
as that used in M06, but we
acknowledge that the choice of priors is arbitrary to a degree. Whilst
changing the prior range, i.e.\ $\Delta w_{0}$ and $\Delta w_{1}$, will
quantitatively affect $\ln B$ calculations it will not qualitatively
change the FoMs we are considering. Furthermore, as discussed in M06,
different (sensible) prior choices will not have a serious impact on
the interpretation of the resulting Bayes factors.

\subsection{P10 correction}

Early in the process of adding the new FoM options to the optimization
we became aware of a coding error that made the original results in
P10 incorrect. We have fixed this error and present the updated
results in Appendix A. The main differences are a reduction in the
DETF FoM by a factor of 3 and that the optimization is qualitatively
unchanged by including curvature as a nuisance parameter. The latter
of these results is found to be also true of the model selection FoMs
we investigate in the following, therefore we do not explicitly
consider curvature in any of our presented findings.

\section{Testing our model selection Figures of Merit}
As mentioned, both the nested $\ln B(-1,0)$ and SDDR area$^{-1}$
computations are quite slow. To deal with this issue,
discrete, i.e.\ manual, optimizations are considered.

Large-scale surveys such as this
are designed with the number of fibres tuned to the required source density,
therefore repeated observations of the same area of sky
are rarely needed. Furthermore the minimum exposure time is nearly always
sufficient to achieve the required S/N on large populations of the
the observed galaxies ($\sim$80\%) which is why the DETF optimization prefers
to maximise the area \cite{par2010}. We therefore chose to set the time and
area to maximum and then manually vary the maximum and minimum redshift
limits. In doing so we have essentially maximised over all other
survey parameters, which allows clearer interpretation of the FoM performance.

\subsection{$\boldsymbol{\ln B(-1,0)}$: an absolute scale for optimization}

\begin{figure}
	\centering
		\includegraphics[width=0.9\columnwidth]{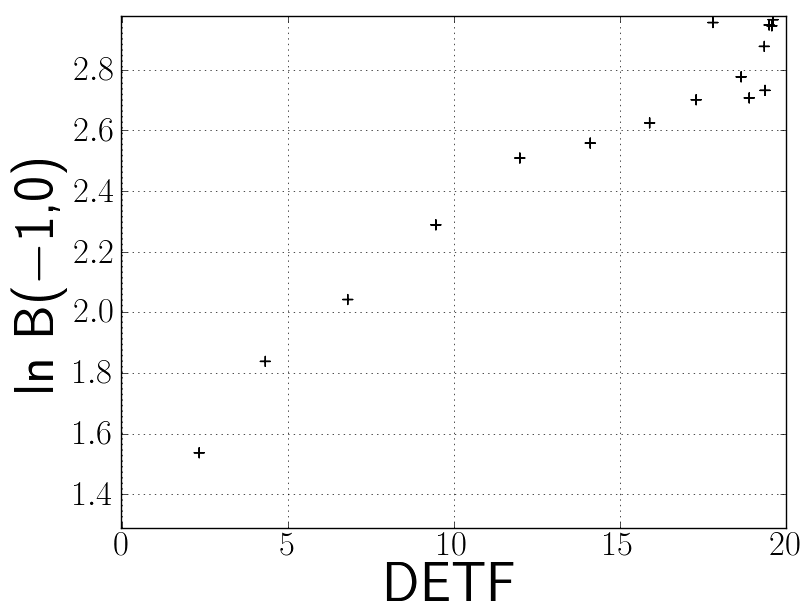}
	\caption{Plot showing the relation between the DETF
          FoM and the nested $\ln B(-1,0)$ FoM. The monotonic relation between the two is clear, with increased scatter seen around the maximal values.}
	\label{fig:NST_DETF2}
\end{figure}

The findings of our nested sampling discrete optimization are
summarised in Figure \ref{fig:NST_DETF2}, which compares the behaviour of
$\ln B(-1,0)$ with that of the DETF FoM.

The mostly monotonic relation we see between the DETF and the nested
$\ln B(-1,0)$ FoMs means that a DETF optimized survey will be extremely similar
to one optimized with $\ln B(-1,0)$. However the latter FoM provides an
absolute scale; in this case $\ln B(-1,0)$ shows the survey is capable of 
strongly preferring
$\Lambda\rm CDM$ when the DETF FoM for the equivalent survey
configuration is still around 60\% of its optimum. This additional
information would be invaluable when deciding on a survey's operating
mode.

\subsection{Gaussian SDDR $\boldsymbol{\ln B(-1,0)}$: an immediately viable model selection
  optimization FoM} 

\begin{figure}
	\centering
        \includegraphics[width=0.9\columnwidth]{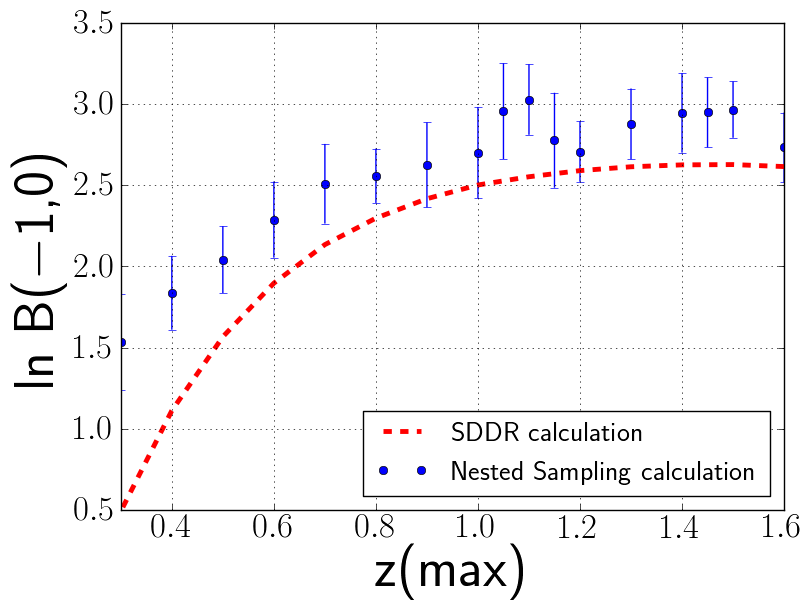}
	\caption{Comparison of the $z_{\rm max}$ dependence of $\ln
          B(-1,0)$ when calculated with nested sampling and Gaussian
          SDDR.  The (blue) circles mark the nested smapling calculations of $\ln B(-1,0)$, whilst the dashed (red) line shows the respective Gaussian SDDR calculations. The Gaussian SDDR numerically underestimates $\ln B(-1,0)$ in a nearly perfectly uniform fashion. We infer that it would be trivial to calibrate it to the more accurate nested calculations. }
	\label{fig:zmax_SDnest}
\end{figure}

By deploying the SDDR Gaussian approximation it is possible to perform full
MCMC optimizations with $\ln B(-1,0)$. However, being an approximation it is
necessary to establish the impact of this simplification on the resulting Bayes
factors. Figure \ref{fig:zmax_SDnest} compares the
Gaussian SDDR calculations with the more accurate nested sampling ones in the case where the upper redshift limit $z_{\rm max}$ is varied.

We can see that the SDDR $\ln B(-1,0)$ typically underestimates the Bayes
factor. While the assumption of Gaussianity has been seen to be good
around the peak of the likelihood \cite{muk2006}, it appears to be less
accurate around the tails. If the information in the tails is
overestimated by this assumption, i.e.\ if in reality the likelihood
falls off more sharply than in the Gaussian approximation, then the
average likelihood and therefore evidence for the evolving dark energy
model will be overestimated. This will result in the underestimation
of the Bayes factor we see here. It would also explain the increased
scatter away from the monotonic relation between the nested
$\ln B(-1,0)$ and the DETF FoM for stronger surveys, clearly
seen in Figure \ref{fig:NST_DETF2}. More accurate surveys will
have tighter likelihood peaks, and therefore any non-Gaussianity
of the tails would be more influential.

\begin{figure}
	\centering
		\includegraphics[width=0.9\columnwidth]{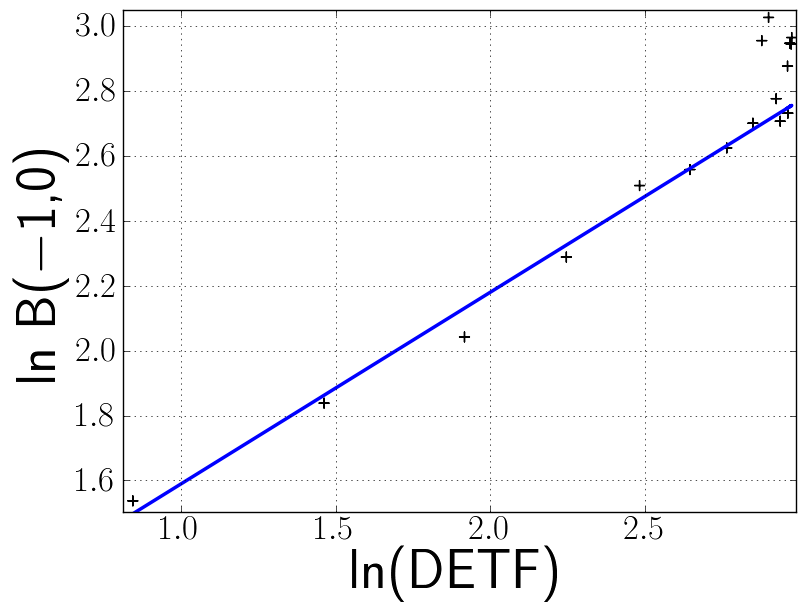}
	\caption{Logarithmic plot of the DETF FoM against the nested
          $\ln B(-1,0)$ FoM. We see that the nested calculation follows the linear relation described by equation \ref{fig:logDETFlnb}.}  
	\label{fig:NST_DETF1}
\end{figure}

Despite this, the general trend is the same, best seen by making a
logarithmic plot of the DETF FoM against nested calculations of $B$,
and writing equation \ref{eq:SDDR2} for the Gaussian SDDR calculation of $B$ in
terms of the DETF FoM:
\begin{equation}
\ln B_{\rm SDDR}(-1,0) = \ln \left(\frac{\Delta w_{0}\Delta
  w_{a}}{2\pi}\right) +  \ln\left(\rm FoM_{DETF}\right) \,.
\label{fig:logDETFlnb}
\end{equation}
Figure \ref{fig:NST_DETF1} shows how the nested calculation of $\ln
B(-1,0)$ follows this linear relation well, despite the increased deviations
around the highest values. This implementation of SDDR presents a good
alternative to the nested sampling approach, and furthermore it is as quick as the
DETF optimization with only a few extra calculations required. Its
underestimation of the Bayes factor is also seen to be almost uniform
across the redshift range of Figure \ref{fig:zmax_SDnest}, and we
therefore infer that it would be simple to calibrate the SDDR FoM to
gain more accurate estimates of $\ln B$ by performing a few nested sampling computations of $\ln B$.

\subsection{Area$\boldsymbol{^{-1}}$: an informative optimization FoM
  with potential for practical application} 

\begin{figure}
	\centering
        \includegraphics[width=0.9\columnwidth]{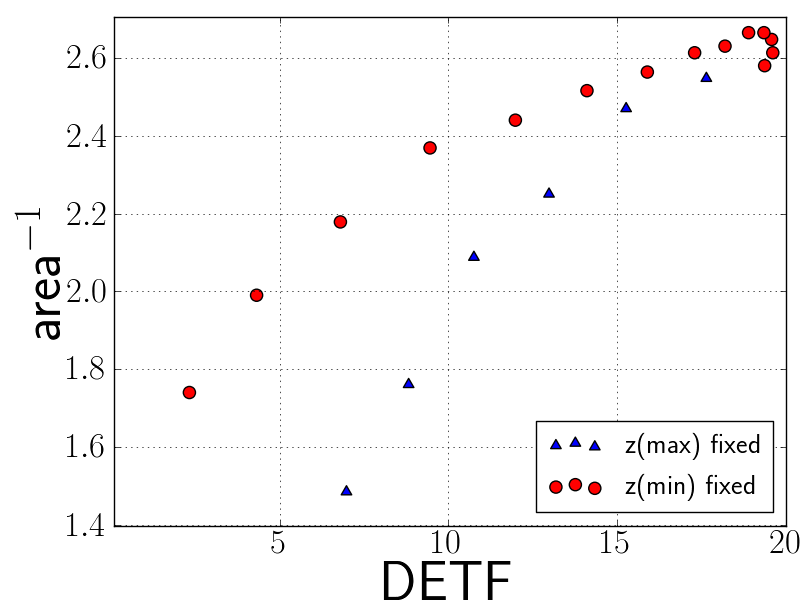}
	\caption{DETF FoM performance compared with that of the
          area$^{-1}$ FoM. The (blue) triangles result from a manual 
          optimization where $z_{\rm min}$ alone is adjusted,
          and the (red) circles
          from adjusting $z_{\rm max}$ only. There is a clear monotonic relation
          between the two, with general agreement around the maximum.
          FoMs were normalised with respect to their maximum.}
	\label{fig:DETFvsAREA}
\end{figure}

The Gaussian SDDR approximation also allows investigation of the
area$\boldsymbol{^{-1}}$ FoM but, as with the nested calculations of $\ln B(-1,0)$,
computational limitations mean that we are again restricted to manual
optimizations. 

Figure \ref{fig:DETFvsAREA} shows the monotonic relation between the
area$^{-1}$ FoM and that of the DETF. This further supports the
widespread adoption of this parameter estimation FoM. The area$^{-1}$
FoM is seen to attain a performance only slightly weaker than its
optimum while the DETF is only 50\% of its optimum. This means that
the model selection ability of a survey is close to optimal for much
weaker configurations then the DETF optimization would deem
acceptable.

\subsection{Fixed galaxy density: a faster, accurate approximation}

\begin{figure}
	\centering
		\includegraphics[width=0.9\columnwidth]{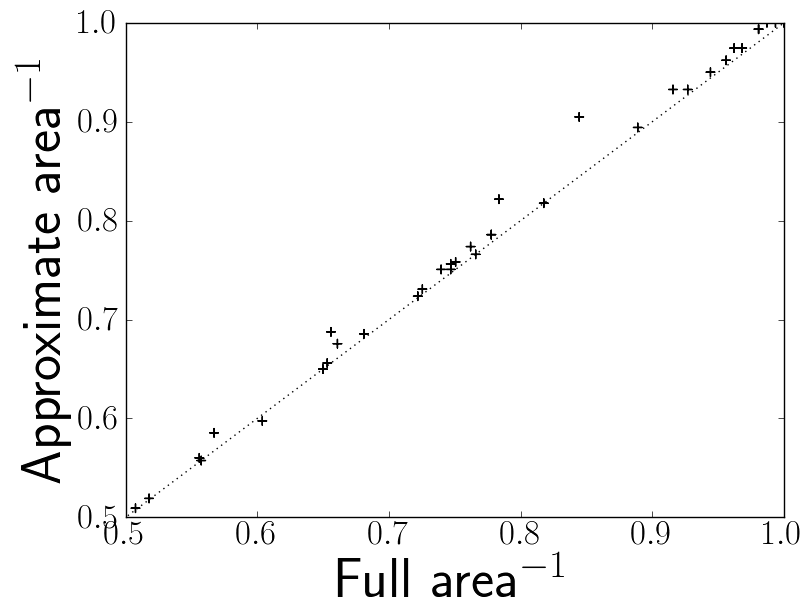}
	\caption{Full implementation area$^{-1}$ FoMs plotted against the
          corresponding values found using the fiducial galaxy approximatation.
          In both cases the FoMs have been normalised with respect to their
          maximum. Clearly the approximation is very good.}
	\label{fig:FULLvsAPPROX}
\end{figure}

As the area$^{-1}$ FoM is slow to compute we consider a further approximation.
We exclude the modelling of the galaxy density, required for the
Seo \& Eisenstein (2007) transfer function, from the $w_0$--$w_a$ gridding. That
is to say we assume that $\Lambda\rm CDM$ is the true model for all
calculations of galaxy density, regardless of the assumed cosmology
in calculating the Bayes factor. In doing so we reduce the number of
times the galaxy density must be estimated from of order 400 per FoM to
1. There are 10,000 FoM calculations in an average optimization, so
this substantially reduces the calculation time.

The results of this are summarised in Figure \ref{fig:FULLvsAPPROX},
showing this to be an extremely good approximation and that this FoM
is not particularly sensitive to galaxy density. This provides a much
faster alternative to the full implementation of the area$^{-1}$ FoM.

\subsection{FoM performance overview}

\begin{figure}
	\centering
		\includegraphics[width=0.9\columnwidth]{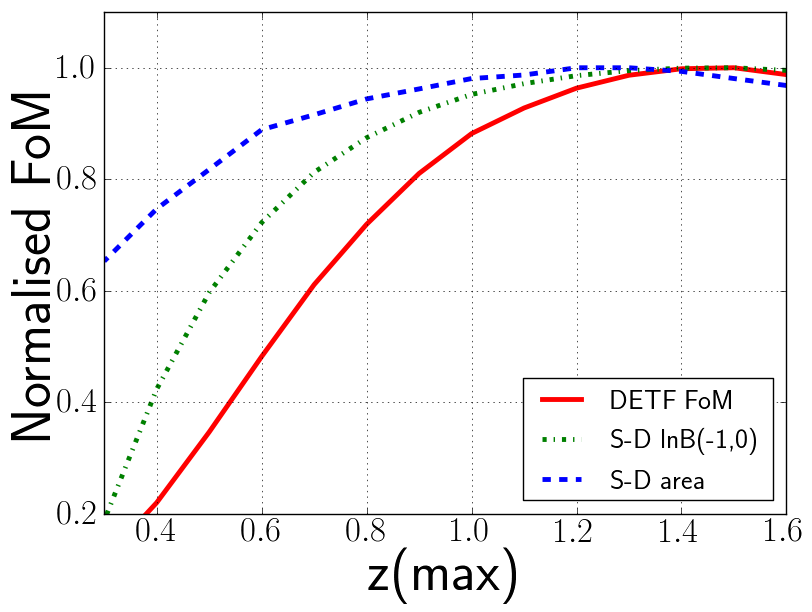} 
	\caption{Optimization results comparing the various FoMs
          investigated. All FoMs were normalised with respect to their
          maximum values. The (blue) dashed line marks the SDDR area$^{-1}$
          FoM, the (green) dot-dashed line plots the SDDR $\ln B(-1,0)$
          FoM, and the (red) solid line shows the DETF FoM.
        }
	\label{fig:FoMcompare}
\end{figure}

Figure \ref{fig:FoMcompare} provides a summary of this investigation
with all FoMs (excluding the nested sampling calculation of $\ln B(-1,0)$) plotted as a function of maximum redshift. From this plot
and that of Figure \ref{fig:zmax_SDnest} we see that the model
selection FoM would deem a maximum redshift of about 0.7 to be
acceptable. Clearly a maximum redshift of between 1.1 and 1.6 is
optimal using the DETF, but it is not clear how much this upper limit
may be reduced before the survey will not fulfil its desired purpose.

Similarly a survey that extends to higher redshifts than 1.6 will be less than
optimal but will, up to a point, be deemed suitable for its purpose by
both model selection FoMs. This highlights the value of this approach
as our modelled survey has the capability to push to higher redshifts; such high-redshift observations are extremely useful for
ancillary cosmology and astronomy. The model selection optimizations 
provide well informed flexibility of this upper $z$ limit.

The absolute scale of $\ln B(-1,0)$ and the additional information from
area$^{-1}$ are very useful when considering time allocations. For
example, a dark energy survey will have limited time on a telescope;
being able to provide detailed information on the required time would
be vital for requesting more time be allocated, or when sharing an
overall allocation between different independent observation modes
within the same project.

As the FoM have all been seen to have a monotonic relation,
information from DETF and $\ln B(-1,0)$ optimizations can be used to perform
similar discrete area$^{-1}$ optimizations as done here, with
negligible expense of time and effort.

\section{Application to spectroscopic BAO surveys}

\subsection{Optimising SuMIRe PFS for model section ability}

We now apply the model selection FoMs that have been investigated so
far to a practical optimization. To make this as relevant as possible
we update the original P10 survey parameters to be as close as can
be to SuMIRe PFS. This involves increasing the mirror diameter from 8m
to 8.2m and the fibre diameter from 1 arcsec to 1.2 arcsecs, while the
target signal-to-noise ratio and throughput are slightly reduced from
the original. The specifications for the survey we optimize in this
section are given in Table \ref{tbl:sumire}; otherwise the survey
details remain unchanged from our original optimization as summarised
in Table \ref{tbl:specs}. For a full description of SuMIRe PFS see
Takada \& Silverman (2010).

\begin{table}
  \centering
  \begin{tabular}{|l|l|}
    \hline
    Parameter & SuMIRe specification\\
    \hline
    mirror diameter & 8.2m \\
    fibre diameter & 1.2 arcsec \\
    aperture & 0.8 \\
    Signal/Noise & 6.5 \\
    Number of fibres & 3000 \\
    Field of View & $9\pi/16$ sq.deg. \\
    \hline
  \end{tabular}
  \caption{The SuMIRE specification used in our calculations.}
  \label{tbl:sumire} 
\end{table}

As the WiggleZ survey is complete and BOSS (Baryon Oscillation
Spectroscopic Survey) is well under way, the SuMIRe optimization must
concentrate on filling the available observational niche, otherwise
some of its allocated time will be lost on unnecessary repetition of
observations. With this in mind we include the forecast data-points
for BOSS and WiggleZ as prior information for this optimization.

\begin{figure}
  \centering
      \includegraphics[width=0.5\columnwidth]{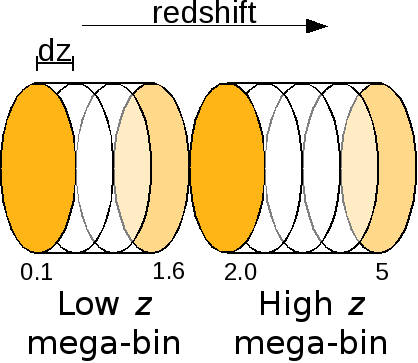}
      \caption{Representation of the $z$ binning method used in
        optimizing SuMIRe PFS.}
      \label{fig:bins2}
  \end{figure}

Although we have so far only considered the redshift range 0.1 to 1.6,
SuMIRe PFS as modelled here is capable of observing redshifts up to
about 4.9. At redshifts greater than 2 the spectrograph can measure
the Lyman-alpha spectral features, however for $1.6 < z < 2$ there
exists an effective blind spot in which no spectral features are
observable by this survey. This optimization therefore considers two
independent redshift mega-bins; the low-redshift mega-bin covering 
$0.1 < z(\rm low) < 1.6$ and the high-redshift mega-bin covering
$2 < z(\rm high) < 4.9$. Figure \ref{fig:bins2} depicts the mega-bin
modelling and Table \ref{tbl:vary2} details the survey variables. The
modelling details of the high redshift mega-bin can be found in P10.

\begin{table}
  \centering
    \begin{tabular}{l|l}
      \hline
      Survey Parameter  & Symbol (mega-bin)\\
      \hline
      Time split between mega-bins & $\tau$(low), $\tau$(high)\\
      Area covered & $A$(low), $A$(high)\\
      Minimum of redshift mega-bin & $z_{\rm min}$(low), $z_{\rm min}$(high) \\
      Maximum of redshift mega-bin & $z_{\rm max}$(low), $z_{\rm max}$(high)\\
      Number of pointings & $n_{\rm p}$(low), $n_{\rm p}$(high) \\
    \end{tabular}
    \caption{Survey parameters varied in the SuMIRe optimization.}
    \label{tbl:vary2}
\end{table}

Two types of optimization are performed. Firstly a full MCMC
optimization using the Savage--Dickey $\ln B(-1,0)$ FoM is executed
using 3 different optimization settings:

\begin{enumerate}
\item Varying all 10 survey parameters listed in Table \ref{tbl:vary2}, i.e.\  $\tau$, $A$, $z_{\rm min}$, $n_{\rm p}$
  and $z_{\rm max}$ over both redshift mega-bins,
\item Focusing all time in the low-redshift mega-bin, i.e.\ varying only $A$(low),
  $z_{\rm min}$(low), $n_{\rm p}$(low) and $z_{\rm max}$(low),
\item Focusing all time in the high-redshift mega-bin, i.e.\ varying only $A$(high),
  $z_{\rm min}$(high), $n_{\rm p}$(high) and $z_{\rm max}$(high).
\end{enumerate}

From optimization (i) we establish the optimum time split between the
low and high mega-bins; the optimum redshift and exposure times for
each are then found from (ii) and (iii). Discrete optimizations can
then be performed with the Savage--Dickey area$^{-1}$ FoM. To do this
we fix the redshift limits and exposure time according to the $\ln B$
optimization results. By manually varying the time (and therefore
area) we can examine this FoM's performance with total time allocation
and its split across the redshift mega-bins.

The SuMIRe project has moved on a great deal from the version modelled
here, for example the current design has no redshift blind spot and
can in principle observe as deep as $z=10$ if targets are available \cite{mur2011}. As such direct
comparison cannot be drawn; we instead use the observational technique
outlined in the 2010 SuMIRe PFS white paper \cite{tak2010a} as our
reference. In this set-up the survey only covers an area of
2000 sq.deg., limited by the projected area that the Hyper Suprime-Cam
(HSC), needed for pre-selecting PFS's target galaxies, will cover
during its operational lifetime. Its redshift range is 0.6 to 1.6,
exposure time is 15 minutes, and therefore total time is roughly 500
hours.

\subsection{Optimal performance for preferring
  $\boldsymbol{\Lambda\mathrm{CDM}}$ achieved with all time spent
  observing redshifts between 0.1 and 1.6} 

\begin{figure}
	\centering
		\includegraphics[width=0.9\columnwidth]{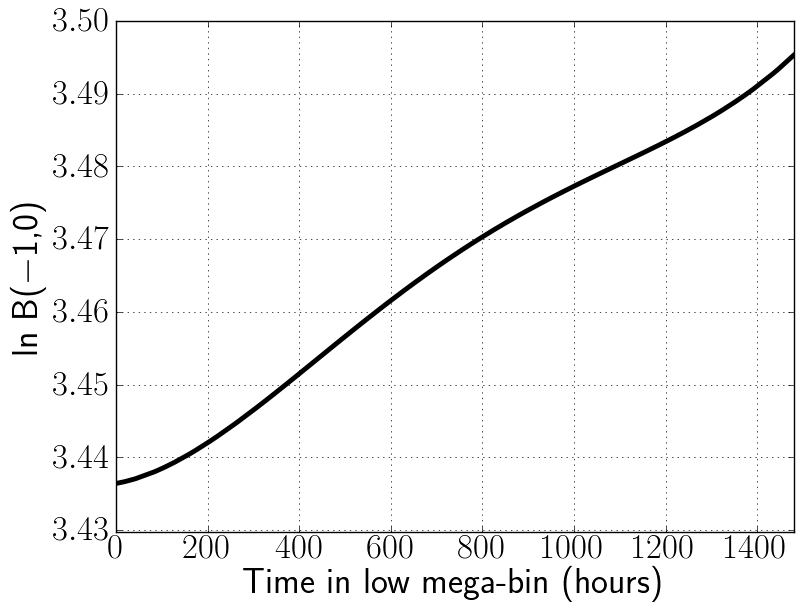}
	\caption{Plot of the time allocation in the low
          mega-bin versus the $\ln B(-1,0)$ FoM; this results from varying all
          parameters in both redshift mega-bins, i.e.\ optimization
          setting (i). Note, any remains of the total
          1500 hours is allocated to the high redshift
          mega-bin. Allocating all 1500 hours to the low mega-bin is
          seen to be preferable}
	\label{fig:timelow}
\end{figure}

The full MCMC optimization using $\ln B(-1,0)$ found that an
improvement in our confidence in $\Lambda\rm CDM$ (if it is the
underlying model) will be gained for a wide range of time allocation
to the low redshift mega-bin, but it is clearly preferable to focus
all time in this mega-bin. This is summarised in Figure
\ref{fig:timelow}, plotting $\tau_{\rm low}$ against $\ln B(-1,0)$.

\begin{figure}
	\centering
		\includegraphics[width=0.9\columnwidth]{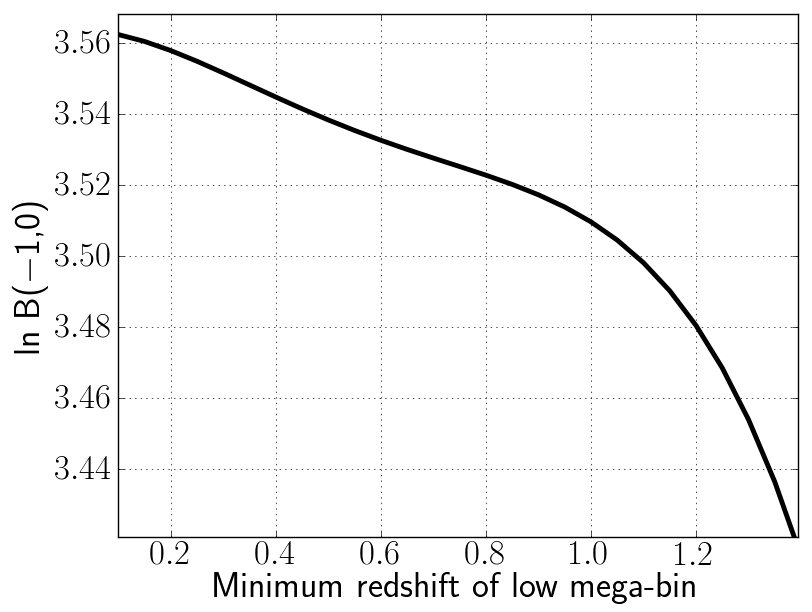}
	\caption{Plot of the minimum redshift of the low
          mega-bin versus $\ln B(-1,0)$ when all time is focused in
          the low mega-bin; this results from varying all
          parameters in both redshift mega-bins, i.e.\ optimization
          setting (i). $z_{\rm min} = 0.1$ is seen to be
          preferable, with model selection performance dropping off
          steeply after $z_{\rm min} = 1.0$}
	\label{fig:zminlow}
\end{figure}

Figure \ref{fig:zminlow} shows how the value of $z_{\rm min}(\rm low)$
affects the survey's ability to prefer $\Lambda\rm CDM$. We find that it
is best to make use of the full redshift range in the low mega-bin,
i.e.\ $0.1 <z <1.6$. However, as long as the lower limit is not greater
than $z=1.0$ there is no great loss of performance.
This has a great deal to do with the fact that the data-points
measured by WiggleZ and BOSS cover the redshift range between 0.1 and
1.0. It also indicates that the reference survey's choice of $z_{\rm
  min} =0.6$ is reasonable.

As with the DETF optimization there is a preference for maximising the
survey volume, and we therefore find that the optimal survey minimises
the exposure time and maximises survey area.

\subsection{For optimal dynamical model selection ability, total time
  of over 1100 hours is preferable, with a minimum of 500 hours spent
  in the low-redshift regime} 
  
\begin{figure}
	\centering
		\includegraphics[width=0.9\columnwidth]{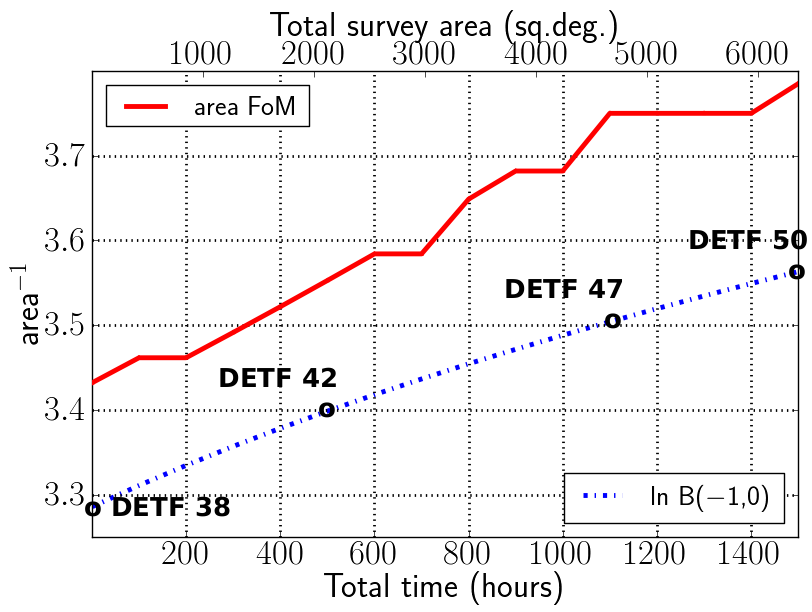}
	\caption{Plot of the total time allocation effect against the
          Area$^{-1}$ FoM; this is achieved by fixing the redshift range
          at 0.1 - 1.6 and exposure time to 15 minutes. We see that
          optimal results can be achieved for time allocations above
          1100 hours.}
	\label{fig:tottime}
\end{figure}

\begin{figure}
	\centering
		\includegraphics[width=0.9\columnwidth]{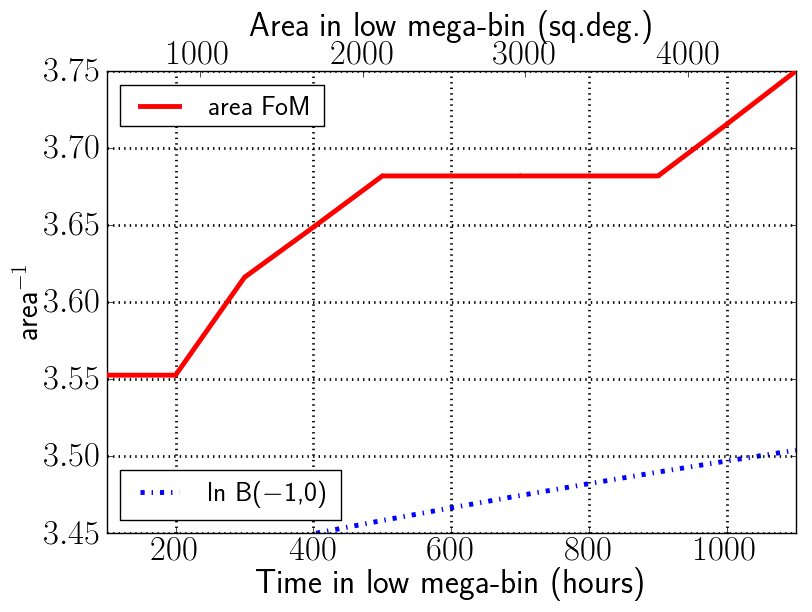}
	\caption{Plot of time
          allocation in the low mega-bin versus the Area$^{-1}$ FoM;
          this is achieved by fixing the redshift range
          at 0.1 - 1.6 and exposure time to 15 minutes. Any
          remains from the total 1100 hours are allocated to the high
          redshift mega-bin. We see that spending over 500 hours in
          the low is preferable for optimal model selection ability. }
	\label{fig:timelow1}
\end{figure}

The time-area$^{-1}$ FoM plots of Figures \ref{fig:tottime} and \ref{fig:timelow1} summarise the findings of our discrete area$^{-1}$ optimization used to investigate the optimal time split.
Their jagged nature is a result of a tipping-point style effect caused by the gridding approach we use for
investigating the dark energy parameter space. This jaggedness is
absent in Figures \ref{fig:DETFvsAREA} and \ref{fig:FoMcompare}
because the range for the FoM is greater by a factor of around 4. An MCMC
style approach to measuring the area$^{-1}$ would produce more gradual
increase of this FoM. The flat-lining should be conceptually smoothed
across, joining the tips of the jags.

Note that in making the total time plot of Figure \ref{fig:tottime},
four different time splits are calculated per total time allocation;
it was found the maximal survey spends all the time in the low mega-bin
regardless of the total time allocated.

The main thing we note is the very limited gain in carrying out this
survey with 500 hours as per our reference survey; $\ln B(-1,0)$
increases from 3.3 to only 3.4, and the prior-space area in which $\Lambda\rm CDM$
cannot be ruled out, i.e.\ area in which $\ln B > -2.5$, reduces from
30\% to only 29\%. This fact is already clear from the DETF FoM with an
increase of only 4. From our model selection optimization a time
allocation of around 1100 hours is best with a minimum of 500 hours
spent in the low mega-bin, but even this promises only a minimal
advance in our knowledge.

SuMIRe PFS as modelled here is only a part of the SuMIRe project; the other
part being the HSC survey, which we've mentioned is used to pre-select
galaxies for PFS. Having the pre-selection survey as an integral part of
the project, operating on the same telescope, is of great benefit in itself,
but the HSC will also be used to measure gravitational lensing. Dark energy
constraints from BAO and gravitational lensing are complementary and as such
their combination dramatically boosts the power of SuMIRe. This is clear from
the white paper's forecasted DETF FoMs which rise from 33
(for both PFS and BOSS surveys), to 217 when the HSC survey is included
(i.e.\ PFS, BOSS and HSC).
Again SDSS and Planck datasets were taken as prior information in calculating
these FoMs \cite{tak2010a}.

It is a pity our modelled survey is not as comprehensive as the full incarnation of
SuMIRe, as the model selection FoM's benefits would be more
transparent; we therefore present this section as a proof of concept
rather than a display of strength. 

\section{Conclusions}

We have discussed the importance of survey optimization, particularly
in finding the appropriate niche for an upcoming survey. The usual
approach of parameter estimation, which maximises the survey's ability
to accurately measure the parameters of interest, assumes that a
particular model is true thereby ignoring the model selection
requirement of these surveys. A future survey's primary aim is to discount
models with the ultimate goal of one to prevail. We therefore attempt to
directly optimize a survey for its intended purpose, model selection.

In doing so we test Bayesian model selection in the context of
optimizing a ground-based spectroscopic baryon acoustic oscillation
survey. To do this we extend an optimization based on the DETF FoM, designing it to instead
target two Bayesian FoMs. For the sake of efficiency we
assume a Gaussian likelihood in both cases. The results of each one's
optimization are compared with that of the original parameter
estimation FoM.

The $\Lambda\rm CDM$ Bayes factor, $\ln B(-1, 0)$, measures the survey's ability to prefer $\Lambda\rm CDM$ if it does transpire to be the correct model. This quantifies the increase in probability of one model over another in light of fresh data, assuming $w_0=-1$ and $w_a=0$ for
all calculations.

For a second FoM, which we call area$^{-1}$, the dark energy parameter space is gridded and discrete
calculations of $\ln B(w_0,w_a)$ are made for each point on the
grid. For each of these calculations the $w_0$, $w_a$ values for that
point in parameter space are assumed. The area in which $\Lambda\rm
CDM$ was not discounted when presumed incorrect, i.e.\ all places where $\ln B(w_0,w_a)>-2.5$,
was calculated. The smaller this area, the more effective a survey is
at model selection and its inverse forms our second FoM.

The $\ln B(-1,0)$ FoM is implemented with only minor adjustment to the original
optimization, and furthermore the calculation time is unchanged. Whilst
this FoM follows the same trend as the original, and
therefore its optimal survey agrees, it does enjoy the added merit of
being an absolute scale allowing interpretation of when the survey is
`good enough'. This added insight is invaluable for making efficient
use of precious survey time or when bidding for extra time
allocations. It also allows for educated flexibility, essential for
adding independent science goals to a project.

The area$^{-1}$ FoM needs some further development to be
useful in full MCMC optimizations, but even as presented here it has
potential for immediate application. We again find this FoM follows
the trends of the original optimization, as such the resulting optimal
surveys will be very close. However there is more information to be
had using this model selection approach; where the other FoMs increase
gradually in a near linear fashion before reaching a brief peak, the
area$^{-1}$ FoM is seen to reach values close to optimum for configurations
the usual approach would deem relatively weak. This approach provides better
insight into the flexibility of the survey's observational strategy.

Whilst these results do not blow the usual parameter estimation
approach out of the water, they do present a powerful
alternative. Considering the extreme simplicity with which the $\ln
B(-1,0)$ FoM can be implemented it seems wasteful to not at the
very least calculate this alongside the DETF FoM. As mentioned the area$^{-1}$ FoM also has
potential for immediate application even before improvements such as
MCMCing the area or parallelising are considered. Whilst there might
not always be a strong trend such as the need to maximise survey
volume to allow the fixing of everything but time, there will
invariably be refined optimizations for which the discrete method used
here is applicable.

Dark energy surveys often require several probes be exploited, some
requiring different observational strategies; for example, the Dark
Energy Survey has one operational mode for taking supernovae data, and
another for everything else \cite{dar2004}. These Bayesian FoMs are perfect for identifying the best time share between such
observational modes, ensuring each independent survey mode is good
enough to achieve its design goals.

The Gaussian approach used here is seen to be reasonable for the $\ln
B(-1,0)$ FoM. We do not however investigate its impact on
the area$^{-1}$ FoM, which requires future work as the
assumption will be less appropriate away from the fiducial point in
dark energy parameter space. However it seems unlikely that the
general trend will be severely altered, and as such development to
improve the speed of our implementation would be beneficial.

Despite the fact we have focused on a particular survey, the methods
discussed here are applicable to any dark energy optimization;
furthermore this can, in theory, be employed beyond dark energy
surveys and even cosmology itself. Therefore it will be interesting to
see how such FoMs fare under different circumstances,
especially in cases where an absolute scale is of particular use, as is
the case for multi-probe surveys.

\section*{Acknowledgements}

C.W.\ was supported by the South East Physics Network (SEPnet),
A.R.L.\ and P.M.\ by the Science and Technology Facilities Council
[grant numbers ST/F002858/1 and ST/I000976/1], A.R.L.\ by a Royal Society--Wolfson Research Merit Award, and D.P.\ by the
Australian Research Council through a Discovery Project grant. We thank Bruce Bassett and Bob Nichol for comments.


\appendix

\section{Revision of P10 results} 

\begin{figure}
  \centering
  \includegraphics[scale=1.2]{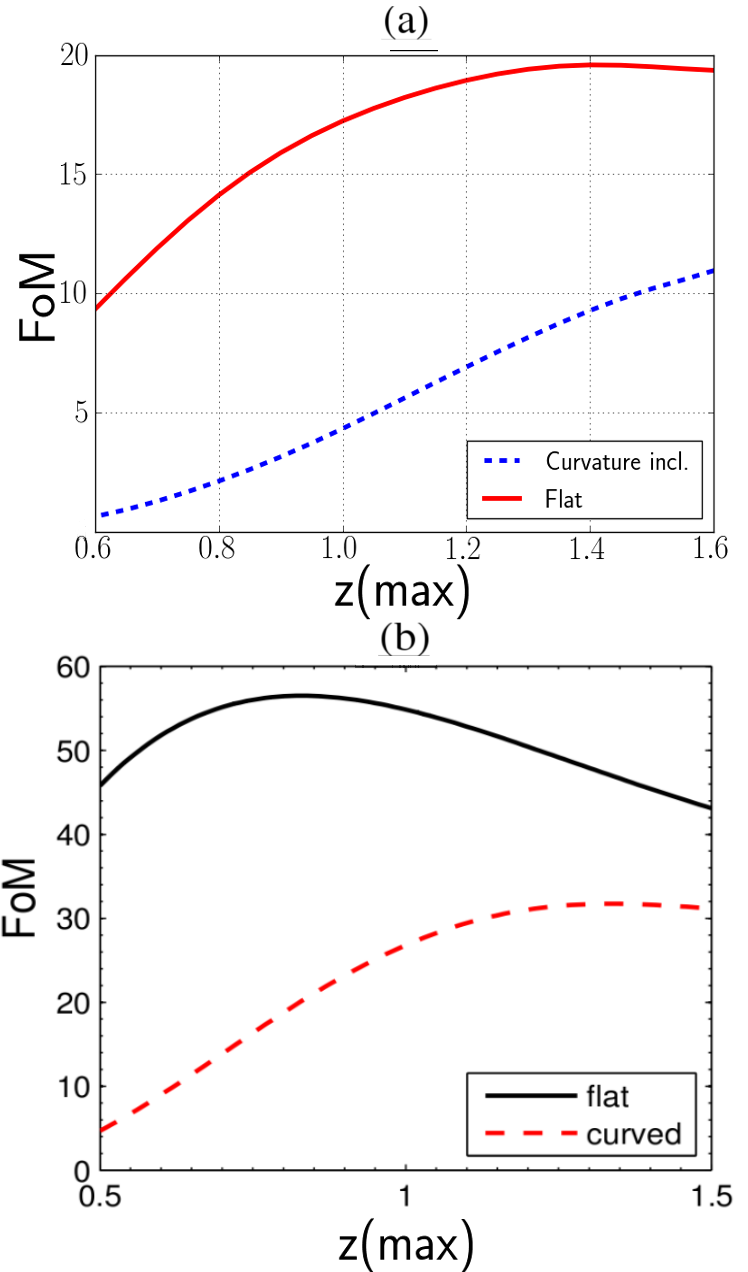}
  \caption{(a) Revised $z_{\rm max}$ - DETF FoM relation when
    the universe is assumed flat (red solid
    line) compared to when curvature is included (blue dashed line)
    (b) Original $z_{\rm max}$ - DETF FoM relation when the universe is
    assumed flat (black solid line) compared to when curvature is
    included (red dashed line). [Lower image from P10.]}
  \label{fig:FlCvCompare}
\end{figure}

The optimal survey parameter values found with the debugged version of
the optimization are summarised in Figure \ref{fig:FlCvCompare} and
Table \ref{tbl:optimresults}. The most pronounced difference compared
with P10 is the reduction of performance predicted by this
optimization, with returned FoMs around a factor of 3 smaller. This
has impact on the forecasted fitness of this survey, which at the time
of P10's publishing performed fairly well alongside its competitors. These results
show it would not have been as strong as previously thought.

\begin{table} 
  \centering
    \begin{tabular}{|l|c|c||c|c|}
      \hline
      \hline
      Survey Parameter & Flat & Curved \\
      \hline
      $A_{\rm low}$(sq.degs)& 6300 &  6300 \\
      $\tau_{\rm low}$ (hours) & 1500 & 1500 \\
      $z_{\rm min} (\rm low)$& 0.1 & 0.1 \\
      $z_{\rm max} (\rm low)$ & 1.5 & 1.6 \\
      exposure time (mins) & 15.0 & 15.0\\
      number density (low) $\rm (h^{3}/Mpc^{3})$  & $5.2 \times 10^{-4}$ & $4.4 \times 10^{-4}$  \\
      number of galaxies (low)  & $4.8 \times 10^{8}$  & $4.6 \times 10^{8}$ \\
      \hline
      FoM  & 20 & 11 \\
      \hline
      Unoptimized FoM  & 7 & 3 \\
      \hline
    \end{tabular}
    \caption{Revised optimal survey parameters obtained with
      the debugged version of the WFMOS optimization.}
    \label{tbl:optimresults}
\end{table}

The improvement in performance achieved by carrying out optimization
is around a factor of 3 in both the original and debugged results,
this is seen by comparison with the unoptimized FoMs. There is also
general agreement that including curvature degrades the FoM, due to
presence of an extra parameter in the Fisher matrix diluting
constraining power on the dark energy parameters.

In the original results, it was inferred that a gamble exists in the
assumption of flatness. P10 found that to proceed with the flat
optimization settings, i.e.\ limiting observation to $0.1 < z < 0.7$,
would be  damaging if curvature does in fact require
constraining. However the new results find far less of a clash of
interests, with the inclusion of curvature pushing the upper redshift
limit up from 1.5 to only 1.6. This is best quantified in Table
\ref{tbl:cvecompare} where the two different optimal survey
configurations are tested under the opposite cosmological
assumption. This is achieved by calculating the FoM with curvature
included, with the survey configuration fixed according to the results
of the flat optimization, and vice versa.

\begin{table} 
  \centering
    \begin{tabular}{|l|c|c||c|c|}
      \hline
      \hline
        & Flat opt. &  Curv opt. \\
        & $z_{\rm range}=0.1$--$1.5$ & $z_{\rm range}=0.1$--$1.6$ \\
      \hline
      FoM (assuming flat) & 20 & 18 \\
      FoM (Curvature incl.)&10&11 \\
      \hline
    \end{tabular}
    \caption{Revised results of degradation caused by either not
      accounting for curvature in optimization when it is necessary or
      allowing for it when it is not. For all configurations the area
      is set to 6300 and the time in the low to 1500.}
    \label{tbl:cvecompare}
\end{table}

 Clearly there is little difference in the performance of the two
 different optimization results if curvature does need to be
 considered, whereas in the previous result, the flat optimal survey
 achieved a FoM about 50\% poorer than that of the curvature included
 optimal survey.
 
\begin{figure}
  \centering
    \subfigure[$z_{\rm min} (\rm high)$]{\label{fig:zminhigh}\includegraphics[scale=0.35]{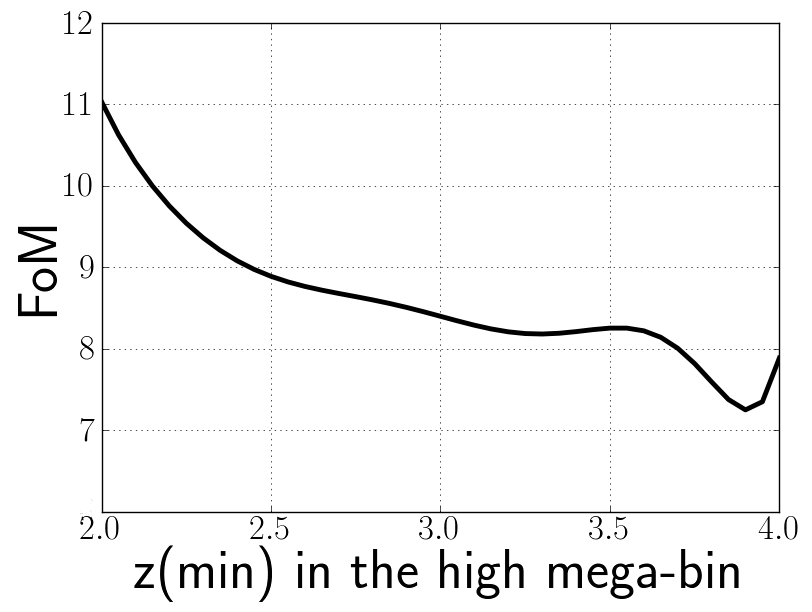}}
    \subfigure[$z_{\rm max} (\rm high)$]{\label{fig:edge-b}\includegraphics[scale=0.35]{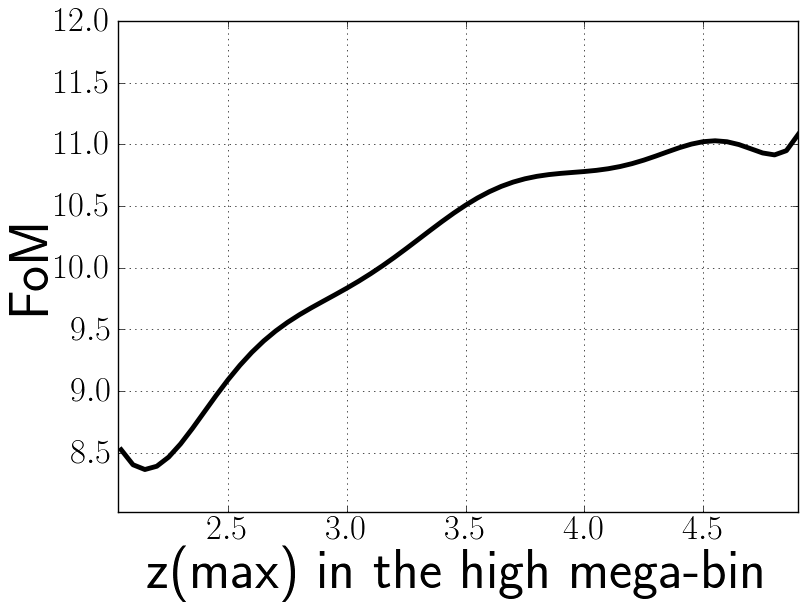}}
  \caption{The $z$ range sensitivity when 380 hours assigned to high-redshift
    mega-bin. (a) Shows how the DETF FoM varies with $z_{\rm min} (\rm high)$,
    and (b) shows how the DETF FoM varies with $z_{\rm max}(\rm high)$.  The
    lack of sensitivity indicates that time spent in the upper mega-bin is
    not vital, but is also not damaging providing the upper redshift limit
    is greater than $\sim$3.6 and the lower 2.0.}
  \label{fig:zCVEhigh}
\end{figure}

 It is worth noting that although the curvature-included optimal
 survey spends all time in the low mega-bin, it was found that there
 is no loss in spending up to 380 hours in the upper redshift band,
 providing that the lower limit is 2.0 and the upper no less that
 3.5. This can be seen in Figure \ref{fig:zCVEhigh} which shows the
 dependence of the FoM on $z_{\rm min}$ and $z_{\rm max}$ in the high
 redshift mega-bin. Furthermore when this survey configuration is
 tested with flatness assumed, the FoM is 17, so there is still no
 serious impact on the power of the survey if considering curvature
 transpires to be unnecessary. This is a positive feature, as it is
 likely that non dark energy science would benefit greatly from such
 time allocation; its presence could have potentially increased
 support for this project.

\bsp

\end{document}